# Normal and inverse magnetocaloric effects in structurally disordered Laves phase $Y_{1-x}Gd_xCo_2$ ($0 \leq x \leq 1$) compounds

Natalia Pierunek[1], Zbigniew Śniadecki[1*], Mirosław Werwiński[1], Bartosz Wasilewski[1], Victorino Franco[2], Bogdan Idzikowski[1]

*corresponding author: sniadecki@ifmpan.poznan.pl

[1]*Institute of Molecular Physics, Polish Academy of Sciences, M. Smoluchowskiego 17, 60-179 Poznań, Poland*
[2]*Condensed Matter Physics Department, Seville University, 41-080 Seville, Spain*

**Abstract**

Magnetic and magnetocaloric properties of $Y_{1-x}Gd_xCo_2$ compounds, where $x$ = 0.2, 0.4, 0.6, 0.8 and 1.0, were investigated experimentally and theoretically. Crystal structures were characterized by X-ray diffraction (Rietveld analysis) and investigated samples possess the $MgCu_2$-type single phase with $Fd$-$3m$ space group. Melt-spinning process introduced a chemical and topological disorder, which directly affected the magnetic properties. Refrigerant capacity (*RC*), strictly connected to the full width at half maximum $\delta T_{FWHM}$ of the $\Delta S_M(T)$ curve and the maximum of magnetic entropy changes $\Delta S_{Mpk}(T,\Delta H)$, increases from 29 to 148 J/kg with replacement of Y by Gd atoms from $x$ = 0.2 to $x$ = 0.8. *RC* and $\delta T_{FWHM}$ indicate the presence of disorder. Temperature dependences of magnetic entropy change $\Delta S_M(T,\Delta H)$ and *RC* were measured in as-quenched and annealed state for $Y_{0.4}Gd_{0.6}Co_2$. This particular composition was chosen for detailed investigation mainly due to its Curie point ($T_C$ = 282 K), which is close to the room temperature. After isothermal annealing ($\tau_a$ = 60 min, $T_a$ = 700°C) *RC* decreased from 122 to 104 J/kg, which clearly indicates the homogenization of the heat treated sample. Furthermore, observed inverse magnetocaloric effect is associated with the presence of antiferromagnetically coupled Gd and Co magnetic moments. The phase transition temperature increases with increasing Gd content from 74 to 407 K for $Y_{0.8}Gd_{0.2}Co_2$ and $GdCo_2$, respectively. Within the FPLO-LDA DFT method the non-magnetic ground state for $YCo_2$ and the magnetic ground state for $GdCo_2$ are predicted in agreement with experiment. The dependence of calculated total and species-resolved magnetic moments on Gd concentration reasonably agrees with available experimental data.

1. Introduction

The magnetocaloric effect (MCE) is described by the adiabatic temperature change $\Delta T_{ad}$ or isothermal magnetic entropy change $\Delta S_M(T,\Delta H)$ under applied magnetic field $\mu_0 H$. The seeking for new magnetocaloric materials is performed aiming at magnetic refrigeration and has been significantly intensified in the recent years [1-3]. In the first order magneto-structural phase transition (FOMT) materials, magnetization $M$ exhibits an abrupt change at the magnetic ordering temperature $T_C$, which is due to the cooperative effect of structural and magnetic transitions leading to a giant magnetocaloric effect (GMCE) [4]. GMCE is characterized by a very narrow and large peak in the temperature dependence of $\Delta S_M(T,\Delta H)$ and by its hysteretic behavior [5]. Most of the magnetic systems investigated in the context of the MCE are crystalline, but in the recent years some emphasis has been put also on the amorphous and nanocrystalline soft magnetic alloys. On the other hand, the second order magnetic phase transition (SOMT) materials (also these with amorphous structure) are characterized by the absence of thermal and magnetic hysteresis, good ability to tune the Curie temperature, low eddy current losses [6] and, in comparison to FOMT materials, exhibit broader $\Delta S_M(T,\Delta H)$ peak, often resulting in enhanced values of the refrigerant capacity (RC). RC parameter is defined as the area under the $\Delta S_M(T,\Delta H)$ curve with temperature interval $\delta T_{FWHM}$ (full width at the half maximum of $\Delta S_M(T,\Delta H)$ peak) as the integration boundaries [7]. Recently, attention has also been paid to crystalline materials with structural disorder introduced for example by rapid quenching. In such systems, stochastic occupation on some crystallographic positions or vacancies are observed. The structural disorder directly affects the MCE parameters, such as magnetic entropy change $\Delta S_M(T,\Delta H)$, maximum value of $\Delta S_M(T,\Delta H)$ defined as $\Delta S_{Mpk}(T,\Delta H)$, and refrigerant capacity. Nowadays, it is of great interest to understand and analyze the mentioned dependences for different disordered systems [8, 9]. Amaral and Amaral [10] have shown a strong influence of disordered structure on magnetocaloric properties. Theoretically, structural disorder broadens the temperature range of magnetic phase transition due to differences in short range order. Therefore, it affects the temperature dependences of magnetization and magnetic entropy change. Moreover, the values of $\Delta S_{Mpk}(T,\Delta H)$ are smaller, but in many cases the broadening of $\Delta S_M(T,\Delta H)$ peaks causes also the increase of $\delta T_{FWHM}$ and RC. In this sense, structurally disordered crystalline systems can be compared to the amorphous alloys. The interplay between the structural disorder and magnetic interactions opens new possibilities for the studies of unique physical phenomena.

RECo$_2$ (RE – rare earth element) Laves phases due to their distinctive properties belong to the group of intensively investigated materials. C15 type Laves phases of MgCu$_2$-type structure (space group *Fd*-3*m*) have been the intermetallic compounds of interest for magnetocalorics for many years [11, 12]. In this type of phases there are two magnetic sublattices: one formed by the atoms of the rare earth elements and the other formed by Co atoms and characterized by itinerant electron magnetism. For Gd-substituted Y$_{1-x}$Gd$_x$Co$_2$ system magnetic sublattices couple antiparallel and the phase transition is of the second order type [13]. There is also strong influence of the concentration of Gd on magnetic properties of Y$_{1-x}$Gd$_x$Co$_2$ ($0 < x \leq 1$) [14]. The YCo$_2$ compound is an exchange enhanced Pauli paramagnet and an itinerant electron metamagnet with a characteristic profile of electronic density of states [15]. The valence band of Y-Co compounds can be described as a combination of wider 4*d* band of Y and narrower 3*d* band of Co [16]. A small addition of magnetic Gd atoms allows to fulfill Stoner criterion and to induce cooperative magnetism in this system by shifting narrow Co 3*d* band in relation to the Fermi energy level. Most importantly, it was shown that the magnetic ordering in the YCo$_2$ system can be induced in many different ways [17]. Rapid quenching with the use of melt-spinning technique was reported as one of them [18]. Also, the surface of YCo$_2$ possesses stable magnetic moments due to the changes in the nearest neighborhood of Co atoms, as shown already for thin films and nanocrystalline samples [19]. Therefore, it is assumed that the same effect can be reached by introducing structural disorder. Moreover, the second terminal composition GdCo$_2$ is an exceptional system in RECo$_2$ family, because Gd (4*f$^7$*) atoms possess a small orbital moment and consequently all the effects connected with the spin-orbit coupling are less pronounced.

We report the magnetic and especially magnetocaloric properties of Y$_{1-x}$Gd$_x$Co$_2$ ($0 \leq x \leq 1$) alloys. Studies by *ab initio* calculations were performed for the whole range of concentrations, whereas the experimental analysis was done for Gd concentration from 0.2 to 1.0, due to the fact that the YCo$_2$ system was investigated and reported earlier [18]. Comparison of the experimental results for chemically and topologically disordered system and for the homogenized and equilibrated analogue is provided for Y$_{0.4}$Gd$_{0.6}$Co$_2$ composition - the sample with Curie point close to room temperature. In the second part of paper particular attention is given to the numerical calculations of $\Delta S_M(T, \Delta H)$ and to the scaling behavior analysis [20].

## 2. Experimental and computational details

The polycrystalline master alloys of $Y_{1-x}Gd_xCo_2$ ($x$ = 0.2, 0.4, 0.6, 0.8, 1) were prepared by arc-melting of high purity elements (3N or more) in argon atmosphere. The weight loss during the melting process was determined to be less than 0.1 wt% for each sample. Master alloys were remelted and subsequently rapidly quenched by melt-spinning technique (wheel surface velocity was equal to 40 m/s). Magnetic measurements were performed on the vibrating sample magnetometer (VSM) option of Quantum Design PPMS. Temperature dependences of magnetization were measured from 2 to 380 K, while $M(H)$ curves were measured from 0 to 5 T in the same temperature range. Magnetization curves were used to calculate $\Delta S_M(T,\Delta H)$ and $RC$, while $\Delta S_M(T,\Delta H)$ dependences were utilized to determine master curves for all investigated alloys. An additional $M(T)$ measurement was carried out in a range from room temperature to 500 K in a homemade susceptometer to determine the phase transition temperature of the $GdCo_2$ compound. Curie temperature values were obtained on the basis of field cooled curves ($\mu_0 H = 0.5$ T) from the inflection method (maximum of $|dM/dT|$ versus $T$).

The *ab initio* calculations using the coherent potential approximation (CPA) [21] were performed to treat the alloying in $Y_{1-x}Gd_xCo_2$. The scalar relativistic calculations were done with the full-potential local-orbital minimum-basis scheme (FPLO-5.00-18) [22]. The calculations were carried out with a 16x16x16 $k$-mesh, the PW92 form of the LSDA exchange-correlation potential [23], and simultaneous convergence criteria $10^{-8}$ Ha for energy and $10^{-6}$ for charge density. Gd 4$f$ electrons were treated as valence, Y (4$s$, 4$p$) and Co (3$s$, 3$p$) were handled as semi core electrons. The CPA calculations were performed with a step of $x$ = 0.2 in accordance with the experimental Gd concentrations. The crystal structure for all considered $Y_{1-x}Gd_xCo_2$ alloys is the cubic C15 Laves phase with *Fd-3m* (nr. 227) space group and following Wyckoff positions: (1/8 1/8 1/8) for Y/Gd and (1/2 1/2 1/2) for Co. The lattice parameters for the terminal cases of $YCo_2$ and $GdCo_2$ were optimized and are equal 6.950 and 6.973 Å, respectively. The underestimation of the lattice parameters in comparison to the experimental values (for $YCo_2$ $a$ = 7.215 Å, for $GdCo_2$ $a$ = 7.250 Å) is assigned to the well-recognized overbinding nature of LDA. The lattice parameters for the intermediate concentrations were interpolated with an arbitrary assumption of their linear behavior.

In addition to the calculations made with FPLO one extra set of results has been obtained with the Munich SPR-KKR package [24, 25]. Using the latter code the paramagnetic solutions of the considered $Y_{1-x}Gd_xCo_2$ alloys have been modeled based on the disordered local moments (DLM) method [26] in which the thermal disorder among the magnetic moments is depicted using the CPA. The calculations within SPR-KKR were carried out with a 36x36x36 *k*-mesh, 40 energy points, $10^{-6}$ Ry energy convergence criterion, and the VWN form of the LSDA exchange-correlation potential [27]. The same crystallographic parameters were used as for the FPLO calculations.

### 3. Results and discussion

All of the investigated $Y_{1-x}Gd_xCo_2$ ($0 \leq x \leq 1$) samples have the $MgCu_2$-type structure of the cubic Laves phase with *Fd-3m* space group confirmed by X-ray diffraction. A crystallographic analysis of these samples was presented in previous studies [28]. The zero-field-cooled (ZFC), field-cooled (FC) and field-heated (FH) *M(T)* curves measured in a magnetic field equal to 0.5 T for all investigated compounds are shown in Fig. 1. Curie temperatures were determined to be 74, 204, 282, 350 and 407 K for the *x* = 0.2, 0.4, 0.6, 0.8 and 1.0, respectively. This tendency is correlated with the increasing concentration of Gd atoms which carry a high effective magnetic moment (7.94 $\mu_B$ for free $Gd^{3+}$ ion). With increasing content of Gd, an increase of magnetization is also observed. Nakama *et al.* [14] have shown that, for a concentration of Gd atoms in the range from 0 to 0.13, a Kondo-type behavior occurs. Furthermore, magnetic cluster spin glass state has been observed for a range $0.07 \leq x \leq 0.12$. For the samples with low Gd content the *M(T)* curves bifurcate at low temperatures, probably just due to the presence of cluster spin glass state. Such behavior can be connected with the existence of quenched-in structural disorder (due to the specific synthesis method). Śniadecki *et al.* [18] have previously shown that for rapidly quenched system, magnetic ordering can be induced even in $YCo_2$ Pauli paramagnet, where at low temperatures a glassy structure was proposed. The value of magnetization at 150 K for rapidly quenched $YCo_2$ is 0.25 emu/g [18], which is lower than 1.28 emu/g for the analyzed $Y_{0.8}Gd_{0.2}Co_2$ sample. For the latter composition the maximum in *M(T)* is observed in addition to the typical transition from the paramagnetic to ferrimagnetic state which occurs at 74 K (in agreement with results for $YCo_2$ and its itinerant behavior). By analogy, it suggests the presence of a cluster spin glass also in $Y_{0.8}Gd_{0.2}Co_2$. Such a state is usually caused by chemical disorder and formation of clusters with RKKY intercluster interactions or magnetic frustrations. Cluster-glass behavior (mictomagnetic ordering) was previously shown also for $DyMn_{6-x}Ge_{6-x}Fe_xAl_x$ ($0 \leq x \leq 6$), where a noticeable bifurcation of ZFC and FH/FC curves was observed at low temperatures [29]. $M(T)$ dependence for $Y_{0.8}Gd_{0.2}Co_2$ (Fig. 1) is concave downward at higher temperatures, which is related to the presence of Pauli paramagnetic regions in a considerable volume fraction of the sample. In a typical exchange-enhanced Pauli paramagnet, $YCo_2$ for instance, a broad peak is observed with the maximum between 50 and 200 K, while in the present case the low-temperature part is obscured by the magnetically ordered state. Typically in such systems, Gd atoms induce long range ordering in Co sublattice, which couples antiferromagnetically to Gd and is very sensitive even for the smallest changes of molecular field.

The strengthening of magnetic interactions is observed also for the present series of alloys. Magnetization irreversibility (between ZFC and FH/FC curves) can also be observed for the $x = 0.4$ sample and suggests the presence of magnetic clusters. This behavior cannot be neglected even for the samples with high Gd content, but a significant magnetic moment can obscure dynamical effects, such as the mictomagnetic behavior mentioned above. For all of the analyzed samples magnetization drops rapidly at the $T_C$ indicating a transformation from ferrimagnetic to paramagnetic state. Tsuchida *et al.* [30] have shown that the magnetic properties of $Gd_{1-a}Y_aCo_2$ ($0 < a < 1$) could be explained in a simplified way through the molecular field approximation for two magnetic sublattices. Further analysis of the mentioned quantities was performed on the basis of the temperature dependence of the reciprocal susceptibility $1/\chi$ (Fig. 2). $GdCo_2$ was omitted from these considerations due to its high transition temperature, which exceeded the working range of the measuring device. The $\chi^{-1}(T)$ behavior of $Y_{1-x}Gd_xCo_2$ ($x = 0.4, 0.6, 0.8$) compounds can be described by a Curie-Weiss law for temperatures above $T_C$. $Y_{0.8}Gd_{0.2}Co_2$ is on the verge of being ferrimagnetic and its $\chi^{-1}(T)$ dependence does not follow the Curie-Weiss behavior. Calculations were performed on the basis of FC curves. Paramagnetic Curie temperatures ($\Theta_p$) are equal to 197, 278, 345 K for $x = 0.4, 0.6$ and $0.8$, respectively and they are in good agreement with Curie temperatures obtained from the inflection method. With increasing concentration of the $Gd^{3+}$ ions the effective magnetic moment per formula unit is slightly decreasing from 0.40 through 0.39 down to 0.37 $\mu_B$/atom for $Y_{0.6}Gd_{0.4}Co_2$, $Y_{0.4}Gd_{0.6}Co_2$ and $Y_{0.2}Gd_{0.8}Co_2$, respectively. Despite of the increase of Gd concentration, the effective magnetic moment remains unchanged.

*Ab initio* calculations were used as a complementary method to interpret the obtained experimental results. Within the FPLO scalar relativistic LDA approach, the Gd 4$f$ shell is completely filled for the majority and remains empty for the minority spin channel in the $GdCo_2$. The well localized 4$f$ states are located around 5 eV below the Fermi energy level and do not influence significantly the 3$d$-5$d$ valence band. Thus, for the considered $Y_{1-x}Gd_xCo_2$ systems there is no need to treat the 4*f* states as core states or with the LDA + U. As within the utilized FPLO method a combination of CPA and full relativistic effects is not allowed we had to limit the model to the scalar relativistic approach. It omits the spin-orbit coupling, thus the orbital contribution to the magnetic moment cannot be reproduced. This approach is justified by the fact that, according to Hund's rules, Gd has a zero orbital magnetic moment and that Hund's rules are well-preserved in the solid state for lanthanides. Supplementary full relativistic calculations conducted with Munich SPR-KKR code confirm that, giving a value of the orbital magnetic moment on Gd equal to 0.034 $\mu_B$. It also agrees reasonably well with the corresponding value of -0.025 $\mu_B$ calculated by Wu [31].

The total energies of the ordered and disordered magnetic states were calculated for a whole range of Gd concentration of the $Y_{1-x}Gd_xCo_2$ system. The results in the form of energy differences are presented in Fig. 3 and show that for the whole range of considered Gd concentration the ordered magnetic state is energetically more favorable than the disordered/paramagnetic (DLM) state. The energies difference $\Delta E$ is the largest for $GdCo_2$ and decreases almost to zero with decrease of Gd concentration. Since $\Delta E$ describes the energy separation between the magnetically ordered and disordered states it can be interpreted as corresponding to the value of the Curie temperature. The value of $|\Delta E|$ for $GdCo_2$ is equal to 10.5 meV/formula which corresponds to a temperature of about 120 K. Although, this result is far from the experimental $T_C$ for $GdCo_2$, measured here as 407 K, grasping the proper order of magnitude still can be considered as a success of this relatively simple approach. The decrease of $|\Delta E|$ with decrease of Gd concentration is also in agreement with the experimentally observed behavior of $T_C$.

Total and species-resolved spin magnetic moments *m* as functions of Gd concentration *x* in $Y_{1-x}Gd_xCo_2$ are presented in Fig. 4. For the pure $GdCo_2$ the spin magnetic moment of Gd is $m_{Gd} = 7.30$ $\mu_B$ and the one of Co, $m_{Co}$, is oriented antiparallel and equals -0.67 $\mu_B$, leading to $m_{total} = 5.96$ $\mu_B$/formula unit. The calculated $m_{total}$ is overestimated mainly due to strong underestimation of $m_{Co}$ and lack of orbital contribution on Co of about -0.1 $\mu_B$/atom. Underestimation of $m_{Co}$ is related to the use of the Perdew-Wang PW92 exchange-correlation potential, while Wu [31] obtained with the von Barth-Hedin formula the value of $m_{Co}$ much closer to the experimental one and equal to -1.24 $\mu_B$. A similar problem of underestimation of $m_{Co}$ in $Co_2B$ has been addressed before with some success within dynamical mean field approximation DMFT [32]. The experimentally determined moment on Co is also antiparallel to the moment on Gd and is estimated to be of about -1.0 $\mu_B$ [30]. Calculated spin moment on Gd reaches a shallow maximum for $x = 0.8$ and the moment of Y ions $m_Y$ decreases monotonically, starting from $x = 0.2$. The critical Gd concentration, which induces magnetic transition, is estimated from the linear part of the total magnetic moment $m_{total}(x)$ dependence (Fig. 4), and equals $x_C = 0.148$, while the AC susceptibility measurements at T = 27 K indicate long-range magnetic ordering from $x_C = 0.14$ [33]. The theoretical values of $m_{total}$ were set together with experimental results in Fig. 4. The magnetization curves were not measured in the saturating magnetic field, which is the main origin of the observed divergence (besides the mentioned theoretical underestimation of $m_{Co}$).

The temperature dependence of the magnetic entropy change $\Delta S_M(T,\Delta H)$ from 2 to 380 K with $\Delta H$ equal to 5 T is shown in Fig. 5. $\Delta S_M(T,\Delta H)$ values were calculated from the Maxwell relation [34]:

$$\Delta S_M(T,\Delta H) = -\mu_0 \int_0^H \left(\frac{\partial M}{\partial T}\right)_H dH,$$

where $\mu_0$ is the permeability of free space. The maximum value of magnetic entropy change, $\Delta S_{Mpk}(T,\Delta H)$, increases with increasing Gd content. $\Delta S_{Mpk}(T,\Delta H)$ is equal to 0.65, 1.62, 2.12 and 2.52 J/kgK for $x = 0.2, 0.4, 0.6$ and 0.8, respectively. $GdCo_2$ was not characterized due to its too high magnetic transition temperature, which exceeded the working temperature of the VSM device. Some features of the curves suggest the presence of a topological and chemical disorder in the investigated samples. $\Delta S_M(T,\Delta H)$ curves have a broad plateau below the main peak, associated with the presence of magnetically inequivalent regions with contributions from the Co clusters as the most probable reason. One can expect non-equivalent positions of atoms in the unit cell and occurrence of imperfections, *e.g.* vacancies and free volumes generated by rapid quenching, as this technique freezes the atoms in metastable positions. $\Delta S_M(T,\Delta H)$ has a maximum around $T_C$ for each compound. With increasing Gd content, $\Delta S_{Mpk}(T,\Delta H)$ shifts to higher temperatures, showing the same tendency as the Curie temperature. The main parameter connected with the width of the $\Delta S_M(T,\Delta H)$ peak is *RC*. The values of refrigerant capacity were calculated from the following relation [34]:

$$RC = \int_{T_1}^{T_2} |\Delta S_M(T,\Delta H)|\, dT,$$

where $T_1$ and $T_2$ are the temperatures of cold and hot reservoirs, which defines the heat that can be transferred between them. *RC* is equal to 29, 66, 122, and 148 J/kg for $Y_{0.8}Gd_{0.2}Co_2$, $Y_{0.6}Gd_{0.4}Co_2$, $Y_{0.4}Gd_{0.6}Co_2$ and $Y_{0.2}Gd_{0.8}Co_2$, respectively. $Y_{0.8}Gd_{0.2}Co_2$ exhibits unique properties for *T* below 45 K, where an inverse MCE is observed for all of the external magnetic fields. In the vicinity of 20 K the value of $\Delta S_{Mpk}(T,\Delta H)$ is the largest and equals 0.16 J/kgK in 5 T. In comparison to $Y_{0.2}Gd_{0.8}Co_2$ ($T_C = 350$ K, $RC_{0-1\,T} = 24$ J/kg, $RC_{0-1.5\,T} = 33$ J/kg), RC in $DyCo_2$ ($T_C = 142$ K) reaches 46 J/kg for the applied magnetic field of 1 T [35].

$GdCo_2$ in homogeneous state (without structural disorder) yields 21 J/kg in a magnetic field of 1.5 T [36].

Values of $\delta T_{FWHM}$ are higher for the present samples than for the similar equilibrated crystalline compounds, *i.e.* a broadening of the magnetic entropy peak is observed. As example, for a fully crystalline Tb-Dy-Co ternary system $\delta T_{FWHM}$ is in the range from 5 to 20 K [37]. Investigated samples show a higher value of $\delta T_{FWHM}$, which is caused by the chemical and topological disorder. The full width at half maximum of the magnetic entropy change peak increases with increasing Gd content and is equal to 57, 58, 72 and 73 K for $x$ = 0.2, 0.4, 0.6 and 0.8, respectively. Differential scanning calorimetry measurements do not show any signs of thermally activated processes, due to the low enthalpy and broad temperature range where the homogenization process takes place. This is rather common and it was encountered even for the transition between two crystalline phases (cubic and hexagonal) in the melt-spun $DyCu_5$ alloy [38]. Isothermal annealing was conducted in order to show the influence of quenched-in disorder. $Y_{0.4}Gd_{0.6}Co_2$ sample was annealed at 700°C for 60 minutes to relax the crystalline structure. Smoothed curves with roughly the same values of $\Delta S_{Mpk}(T,\Delta H)$ = 2.11 J/kgK for the as-quenched and $\Delta S_{Mpk}(T,\Delta H)$ = 2.15 J/kgK for the annealed one were observed. Such behavior was already reported by Singh *et al.* [8]. What is the most important, *RC* decreases by about 13% and $\delta T_{FWHM}$ also changes its value and is equal to 72 K for the as-quenched state and 63 K for the homogenized sample. At low temperatures, negative changes of magnetic entropy are less distinct for the annealed sample, which suggests a partial decay of the glassy state through the homogenization and more uniform distribution of atoms. Occurrence of inverse magnetocaloric effect at low temperatures is related to the antiferromagnetic or cluster glass behavior. Such effect was also related earlier to the spin reorientation phenomena and was observed, for example, by Khan *et al.* [39] in $Ho_{1-x}Er_xAl_2$. It was linked with the geometry of 4*f* charge densities of $Ho^{3+}$ and $Er^{3+}$ and strong magnetocrystalline anisotropy of $HoAl_2$ and $ErAl_2$. The presence of spin reorientation and simultaneous occurrence of first order phase transition can be waived in the analysis of the investigated compounds. The magnetic moment on Gd atoms has a small orbital contribution and consequently such strong effects coming from the crystal field are not applicable to these compounds. To conclude this paragraph, structural disorder yields higher values of *RC*, resulting from the broadened shape of magnetic entropy curves, despite lower values of $\Delta S_{Mpk}(T,\Delta H)$.

The universal curve (master curve) proposed in 2006 by Franco *et al.* [20] and applied to a series of Laves phases in 2010 by Bonilla *et al.* [11] was used to describe phase transitions and to characterize magnetic behavior of the investigated systems. Master curve is described by the formula [20]:

$$\Theta = \begin{cases} -(T-T_C)/(T_{r1}-T_C) & T \leq T_C \\ (T-T_C)/(T_{r2}-T_C) & T \geq T_C \end{cases},$$

where $T_{r1}$ and $T_{r2}$ are the temperatures of the two reference points of each $\Delta S_M(T,\Delta H)$ curve and were selected as those corresponding to 0.5 $\Delta S_{Mpk}(T,\Delta H)$, $T_C$ is the phase transition temperature and $\Theta$ is the rescaled temperature. Universal curve can be obtained by plotting the normalized $\Delta S_M(T,\Delta H)$ data *versus* $\Theta$, which is divided into two ranges: below and above the Curie temperature. For $Y_{0.8}Gd_{0.2}Co_2$ the master curves do not collapse on each other (Fig. 6) and the most significant divergence is observed for $\Theta < -1.5$ and $\Theta > 1.7$. The collapse is relatively good just below and above $T_C$ for $\Theta$ between -1.5 and 1.7. Differences are clearly visible for low $\Theta$ values which correspond to low temperatures, where the inverse MCE and disordered structure are mostly pronounced. Keeping in mind that the system is on the verge of a long-range magnetic ordering and that the sample is still magnetically inhomogeneous, the master curves do not collapse even for the temperatures above $T_C$. Additionally, an increase of the normalized magnetic entropy changes ($\Delta S_M'(\Theta)$) at higher temperatures ($\Theta$ greater than 6) is observed. This may be caused by the ordering of a different magnetic phase. For other investigated compounds the universal curves show similar behavior with one difference. $Y_{0.6}Gd_{0.4}Co_2$, $Y_{0.4}Gd_{0.6}Co_2$ and $Y_{0.2}Gd_{0.8}Co_2$ exhibit a collapse of the master curves on the paramagnetic side for $0 < \Theta < 1.8$. For the ferrimagnetic region, $\Theta$ below -0.2, master curves are separated (stronger divergence for a higher Gd content). For a wider temperature range the separation is much larger, which is clearly visible for the mentioned samples. It may be due to the presence of the chemical and topological disorder, causing magnetic dynamical processes in the investigated compounds and a strengthening of the antiferromagnetic/RKKY couplings with increasing Gd content. It was reported previously by us [28] that in AC susceptibility measurements an additional peak is observed below $T_C$. Its position appears to be frequency dependent and more detailed studies indicated the presence of magnetic dynamic behavior induced by topological/chemical disorder [28]. Position of low temperature peak was rather stable with simultaneous strong increase of $T_C$ and temperature span between both peaks with changing Gd content. Rescaled magnetic entropy change curves for applied magnetic field of 5 T are shown in Fig. 7. $T_C$ for all presented compounds is fixed at $\Theta = 0$. The temperature span between the main peak at $T_C$ and low temperature smeared feature connected with dynamic behavior is increasing with Gd content, as it was observed in [28]. It confirms that the dispersion of master curves is not stochastic and originates from magnetic ordering. As it can be observed in Fig. 6., various curves of the normalized magnetic entropy change *versus* temperature $\Theta$ for different external magnetic fields do not collapse into a single curve for $\Theta < 0$ even in the high magnetic fields. For $\Theta > 0$ the Curie-Weiss law was preserved and the behavior is typically paramagnetic, except for $Y_{0.8}Gd_{0.2}Co_2$, where additional high temperature enhancement of magnetic moment was observed. Inoue and Shimizu [13] have shown by theoretical studies that all compounds from the analyzed series (except the $YCo_2$ Pauli paramagnet) undergo a second order phase transition. On the basis of our research, occurrence of second order phase transitions is suggested despite the dispersion of the master curves.

Summarizing, the most important factors which alter the magnetic and magnetocaloric properties of the samples investigated are: (i) quenched-in structural (chemical and topological) disorder, (ii) magnetic non-uniformity with the coexistence of ferrimagnetic and mictomagnetic ordering, (iii) presence of antiferromagnetic/RKKY interactions at low temperatures.

Structural and magnetic phase diagrams are shown in Fig. 8 and Fig. 9 with the emphasis on the parameters describing magnetocaloric properties of the investigated samples. Diversity of magnetic phenomena is shown, as the presence of paramagnetism and antiferromagnetic/ferromagnetic coupling is documented on the basis of present studies and numerous references. It should be emphasized that the presence of glassy behavior at low temperatures does not exclude simultaneous existence of ferrimagnetically ordered regions. It is clearly visible, that lattice parameter and Curie temperature are increasing almost linearly with increasing Gd content (Fig. 8). As shown in Fig. 9, values of $\Delta S_M(T,\Delta H)$, $RC$ and $\delta T_{FWHM}$ also increase with Gd concentration.

## 4. Conclusions

The magnetic and magnetocaloric properties of $Y_{1-x}Gd_xCo_2$ ($x = 0.2, 0.4, 0.6, 0.8, 1.0$) samples were analyzed. The compounds were prepared by melt-spinning technique, causing chemical and topological disorder. The series of investigated compounds exhibits normal and inverse magnetocaloric effect near the ferrimagnetic-paramagnetic phase transition and at low temperatures, respectively. Low temperature features are derived from the increasing antiferromagnetic coupling (due to the increase of Gd content), along with mictomagnetic behavior (RKKY coupling or presence of magnetic frustrations) [18] [21]. Refrigerant capacity increases with the addition of Gd and is equal to 29, 66, 122 and 148 J/kg, as well as the value of the magnetic entropy peak $\Delta S_{Mpk}(T,\Delta H)$, which is equal to 0.65, 1.62, 2.11 and

2.52 for $x$ = 0.2, 0.4, 0.6 and 0.8, respectively. Curie temperatures of the melt-spun compounds were determined by an inflection method and indirectly confirmed, with high convergence, by magnetic entropy calculations. $T_C$ amounts to 74, 204, 282, 350 and 407 K for $x$ = 0.2, 0.4, 0.6, 0.8 and 1.0, respectively. Magnetic entropy $\Delta S_M(T,\Delta H)$ and master curve $\Delta S_M'(\Theta)$ data clearly show the influence of quenched-in disorder, which alters magnetic properties, contributing to larger values of the refrigerant capacity $RC$ and $\delta T_{FWHM}$ (full width at half maximum of the magnetic entropy change), and a slight decrease of $\Delta S_{Mpk}(T,\Delta H)$. We have demonstrated that the presence of structural disorder significantly broadens the magnetic transition and the temperature dependent magnetic entropy changes. Calculations within scalar relativistic CPA-LDA model for chemically disordered $Y_{1-x}Gd_xCo_2$ systems describe qualitatively its basic magnetic properties. LDA addresses the magnetic ground state of all considered compositions in a proper way. The critical Gd concentration, for which induction of long-range magnetic ordering is observed, is theoretically estimated to be $x_C$ = 0.148. The calculated total magnetic moment *versus* Gd concentration dependence agrees qualitatively with the experiment.

**Figure captions:**

Fig. 1. Temperature dependences of the zero-field-cooled (ZFC), field-cooled (FC) and field-heated (FH) magnetization curves measured at magnetic field $\mu_0 H = 0.5$ T for $Y_{1-x}Gd_xCo_2$ ($0.2 \leq x \leq 1$) compounds.

Fig. 2. Temperature dependence of reciprocal magnetic susceptibility for $Y_{0.8}Gd_{0.2}Co_2$, $Y_{0.6}Gd_{0.4}Co_2$, $Y_{0.4}Gd_{0.6}Co_2$ and $Y_{0.2}Gd_{0.8}Co_2$ ($\mu_0 H = 0.5$ T). Solid line is the fit (Curie-Weiss law) to experimental data.

Fig. 3. Magnetic energy difference $\Delta E$ between ordered and disordered (DLM) magnetic states as a function of $x$ in $Y_{1-x}Gd_xCo_2$ ($0.2 \leq x \leq 1$) as calculated with the Munich SPR-KKR.

Fig. 4 Total and species-resolved spin magnetic moments as a function of $x$ in $Y_{1-x}Gd_xCo_2$ ($0.2 \leq x \leq 1$), calculated with FPLO, treating disorder with the CPA. Squares represent experimental data and circles represent theoretical data in the top left picture.

Fig. 5 Temperature dependences of the magnetic entropy change determined for $Y_{1-x}Gd_xCo_2$ ($x = 0.2, 0.4, 0.6$ and $0.8$) compounds from the magnetization isotherms for a change in the magnetic field from 0 to 5 T.

Fig. 6. Rescaled magnetic entropy change curves for applied magnetic fields ranging from 1 to 5 T with temperature steps $\Delta T = 5$ K for $Y_{1-x}Gd_xCo_2$ ($x = 0.2, 0.4, 0.6$ and $0.8$) compounds.

Fig. 7. Rescaled magnetic entropy change curves for applied magnetic field equal to 5 T with temperature steps $\Delta T = 5$ K for $Y_{1-x}Gd_xCo_2$ ($x = 0.2, 0.4, 0.6$ and $0.8$) compounds.

Fig. 8. Lattice parameter (circles) and Curie temperature (stars) variations in $Y_{1-x}Gd_xCo_2$ ($x = 0.2, 0.4, 0.6$ and $0.8$) compounds with marked following regions: PM – paramagnetic, FiM – ferrimagnetic.

Fig. 9. Phase diagram of magnetocaloric properties (full width at half maximum $\delta T_{FWHM}$ (squares, ticks inside), refrigerant capacity $RC$ (circles, bold ticks outside) and maximum value of magnetic entropy changes $\Delta S_{Mpk}$ (stars)) of $Y_{1-x}Gd_xCo_2$ ($x = 0.2, 0.4, 0.6$ and $0.8$) compounds with defined regions: PM – paramagnetic, FiM – ferromagnetic.

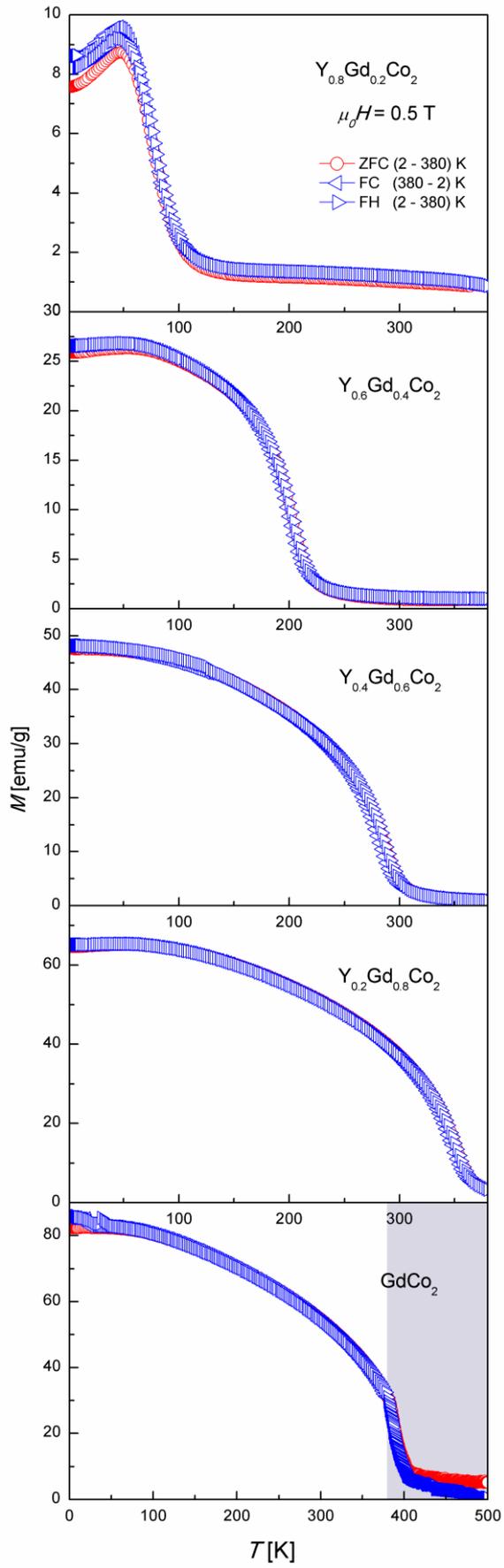

Fig. 1.

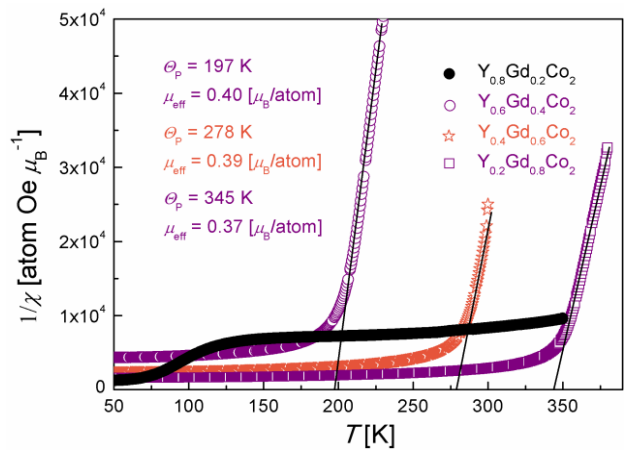

Fig. 2.

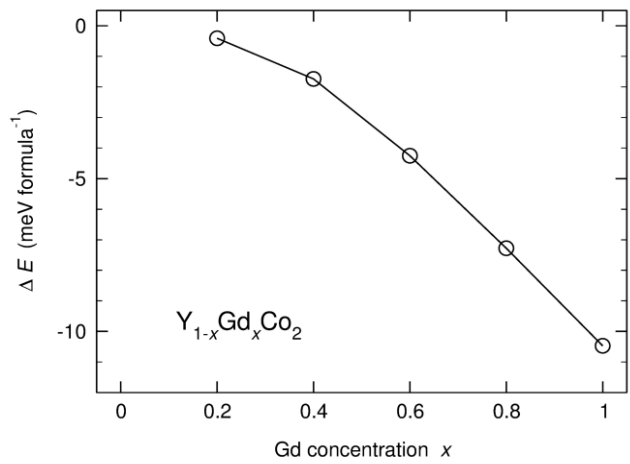

Fig. 3.

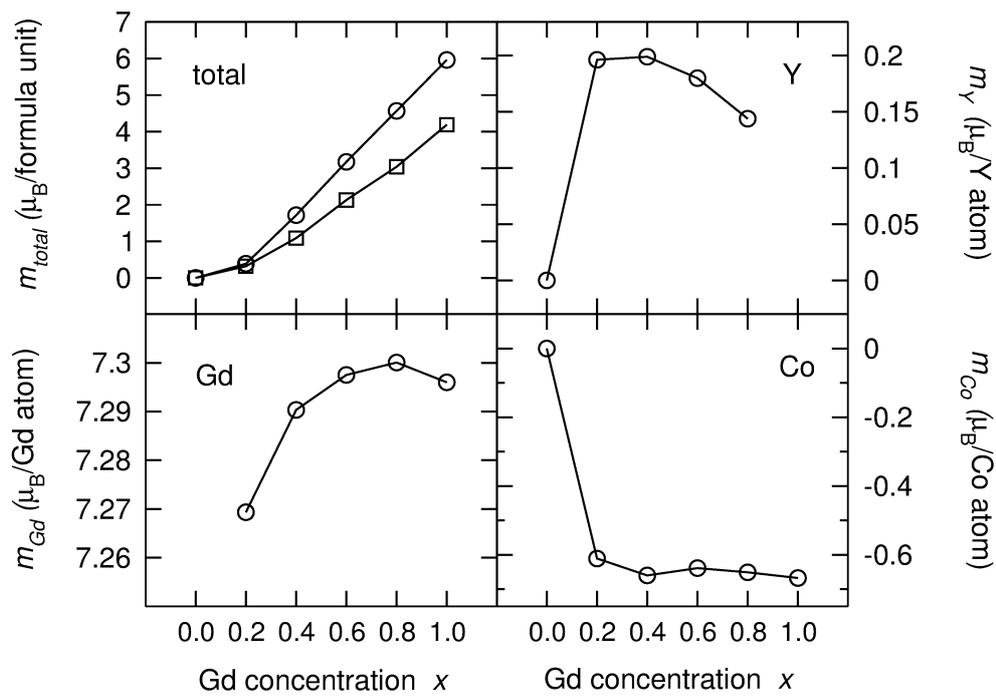

Fig. 4.

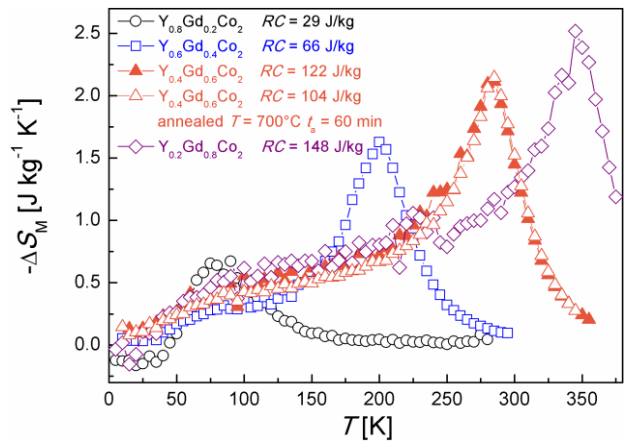

Fig. 5.

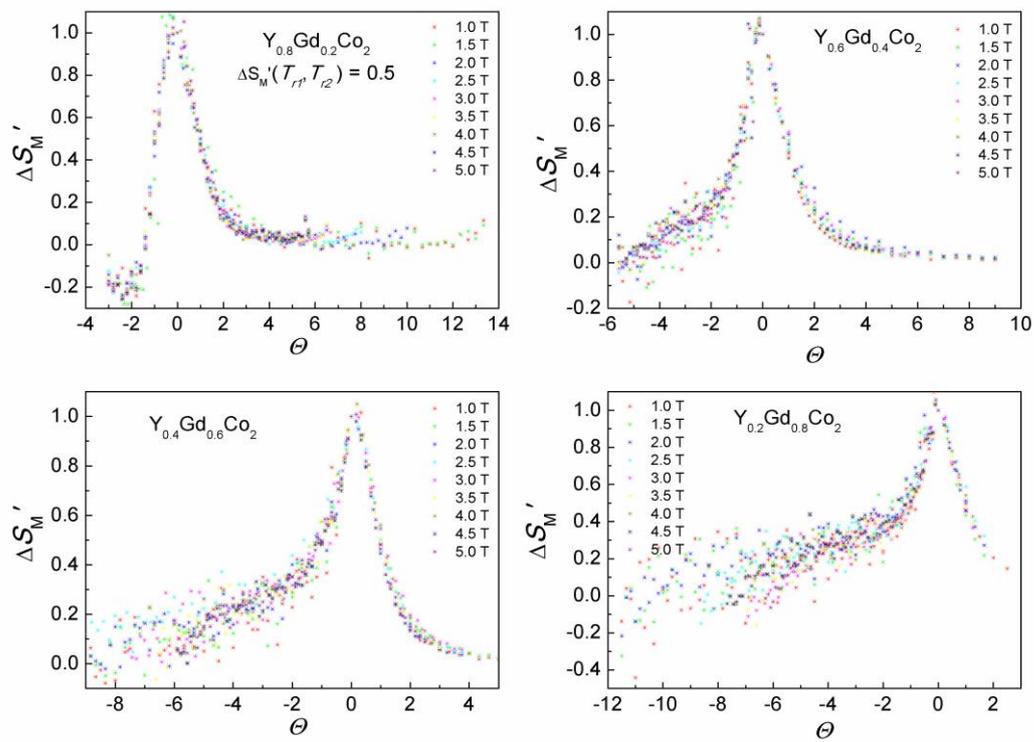

Fig. 6.

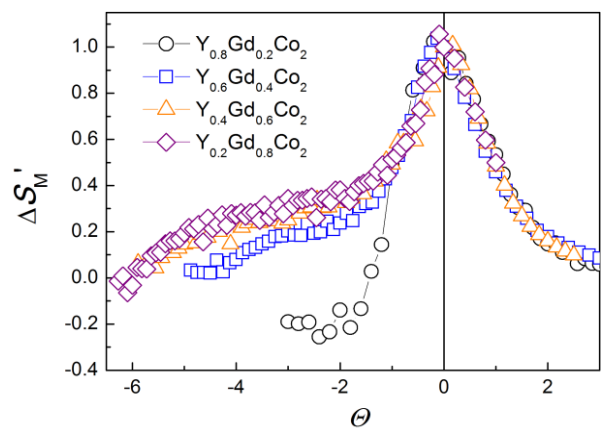

Fig. 7.

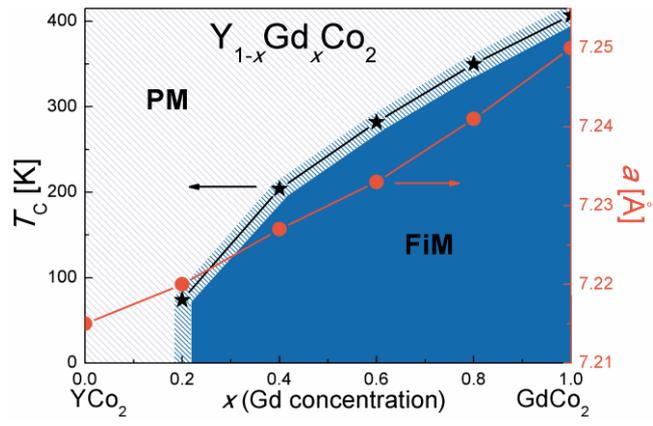

Fig. 8.

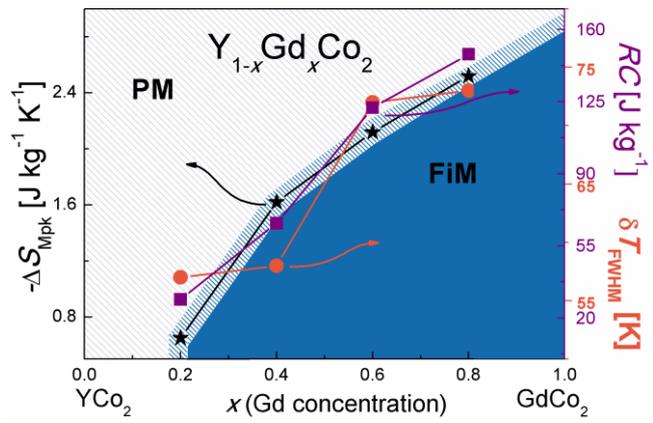

Fig. 9.